\documentclass[letterpaper,twocolumn]{jpsj3}
\usepackage{txfonts}
\usepackage[dvipdfmx]{graphicx}
\usepackage{color}
\usepackage{ulem}

\newcommand{\blue}[1]{\textcolor{black}{#1}}

\title{Magnetovolume Effect on the First-Order Metamagnetic Transition in UTe$_2$}

\author{
Atsushi Miyake$^1$\thanks{miyake@issp.u-tokyo.ac.jp}, 
Masaki Gen$^2$,
Akihiko Ikeda$^3$,
Kazumasa Miyake$^4$,
Yusei Shimizu$^5$, 
Yoshiki J. Sato$^5$, 
Dexin Li$^5$, 
Ai Nakamura$^5$, 
Yoshiya Homma$^5$, 
Fuminori Honda$^{5,6}$, 
Jacques Flouquet$^7$,\\ 
Masashi Tokunaga$^1$, and 
Dai Aoki$^{5, 7}$
}

\inst{
$^1$Institute for Solid State Physics (ISSP), The University of Tokyo, Kashiwa, Chiba 277-8581, Japan\\
$^2$Department of Advanced Materials Science, The University of Tokyo, Kashiwa, Chiba 277-8561, Japan\\
$^3$Department of Engineering Science, University of Electro-Communications, Chofu, Tokyo 182-8585, Japan\\
$^4$Center for Advanced High Magnetic Field Science, Osaka University, Toyonaka, Osaka 560-0043, Japan\\
$^5$Institute for Materials Research, Tohoku University, Oarai, Ibaraki 311-1313, Japan \\
$^6$Central Institute of Radioisotope Science and Safety, Kyushu University, Fukuoka, 819-0395, Japan\\ 
$^7$University Grenoble Alpes, CEA, Grenoble INP, IRIG-PHELIQS, 38000 Grenoble, France
} 

\abst{
The link between the metamagnetic transition and novel spin-triplet superconductivity of UTe$_2$ was discussed thermodynamically through magnetostriction measurements in a pulsed-magnetic field.
We revealed a discontinuous magnetostriction across the metamagnetic transition at $\mu_0H_{\rm m}\approx 35$~T for the applied magnetic fields along the crystallographic $b$ axis in the orthorhombic structure.
The resultanting volume magnetostriction of $\Delta V/V\approx-5.9\times 10^{-4}$ gives the initial pressure dependence of $H_{\rm m}$ by employing the Clausius-Clapeyron's equation, which agrees with previous pressure experiments.
Further, significant anisotropic magnetostriction (AMS), derived by subtracting the averaged linear magnetostriction, was revealed.
Contrary to the weakly field-dependent AMS along the $a$ axis, those along the $b$ and $c$ axes show strong field dependences with a similar magnitude but with opposite signs, indicating its lattice instability.
The relationship between characteristic energy scales of magnetic fields and temperatures was discussed in terms of the Gr\"uneisen parameters compared to the other $f$-electron systems.
The volume shrinkage in UTe$_2$ at $H_{\rm m}$, contrary to the volume expansion in typical heavy fermion metamagnets, pushes to invoke the link with the valence instability related to the itinerant-localized dual nature of the U magnetism. 
}

\begin{document}
\maketitle

\begin{figure}
\begin{center}
\includegraphics[width=0.8\hsize]{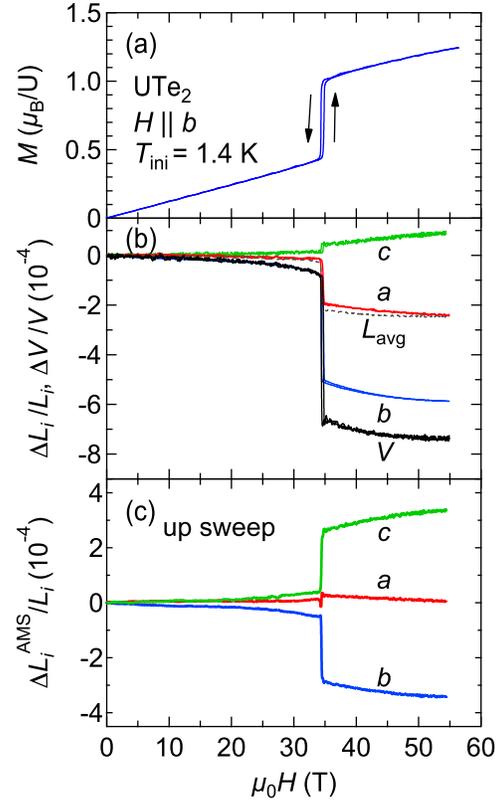}
\end{center}
\caption{(Color online)
Magnetic field dependence of (a)magnetization, (b)linear and volume ($V$) magnetostriction, and (c)anisotropic linear magnetostriction of UTe$_2$ at $T_{\rm ini}$ = 1.4~K for $H~||~b$.
The volume magnetostriction is determined by a summation of the linear magnetostriction, $\Delta V/V = \sum_{i = a, b, c} (\Delta L_i/L_i)$.
$L_{\rm avg}$ is an average of the volume (lattice) magnetostriction determined by $\frac{\Delta V}{3V}$.
The anisotropic magnetostriction is evaluated by $\Delta L_i^{\rm AMS}/L_i=\Delta L_i/L_i-\Delta V/(3V)$.
The $M(H)$ curve is taken from Refs.~\citen{Miyake2019} and \citen{AMiyake2021}.
}
\label{MS}
\end{figure}

Unconventional superconductivity of UTe$_2$ (space group: $Immm$) below $T_{\rm sc}$ = 1.6-2.0~K is a promising candidate for a spin-triplet state \cite{Ran2019,Aoki2019,Rosa2021}. 
Since its discovery few years ago, a number of experimental and theoretical works have been intensively conducted \cite{Aoki_review}.
For instance, puzzling magnetic-field ($H$) effects have been reported from the early stage.
A first-order metamagnetic transition takes place at $\mu_0H_{\rm m} \approx 35$~T, when the field is applied along the $b$ axis \cite{Knafo2019,Miyake2019,Ran2019b,Knafo2021,AMiyake2021} [see Fig.~\ref{MS}(a)].
Notable features appear in its superconducting (SC) phase diagrams, which are significantly sensitive to the field and its applied direction.
For $H~||~b$, $T_{\rm sc}$ decreases to $\sim$15~T, followed by a gradual increase and a sudden suppression at $H_{\rm m}$ \cite{Knebel2019,Ran2019b,Knafo2021}.
More strikingly, another SC phase is induced above $H_{\rm m}$ for a  magnetic  field along specific directions near the [011] direction \cite{Ran2019b, Knafo2021}.
Such reinforcement/reentrant behaviors of SC transition are reminiscent of spin-triplet ferromagnetic (FM) superconductors of URhGe and UCoGe \cite{Levy2005, Levy2007, Aoki2009,Aok12_JPSJ_review, Aoki2019re}.
However, UTe$_2$ does not order magnetically \cite{Sunder2019,Hutanu2020,Paulsen2021}.
The spin fluctuations have been investigated microscopically by NMR \cite{Tokunaga2019} and $\mu$SR \cite{Sunder2019}.
Direct antiferromagnetic (AFM) correlations are detected using inelastic neutron scattering experiments \cite{Duan2020,Knafo2021b}, while FM correlations have not yet been observed experimentally up to date.
These AFM and FM fluctuations may compete for the choice of SC order parameters \cite{Xu2019, Ishizuka2021}.  

Although it is evident that the first-order metamagnetic transition influences the SC phases, the relation between them, particularly the origin of metamagnetism, leaves many mysteries.
For $H~||~b$, a steep increase in magnetization from $M\approx 0.5~\mu_{\rm B}$/U to $M\approx1.0 ~\mu_{\rm B}$/U at $H_{\rm m}$ [see Fig.~\ref{MS}(a)] accompanied by approximately five times enhancement of the resistivity with a clear $H$-hysteresis at low temperatures is observed \cite{Knafo2019,Miyake2019,Knafo2021,AMiyake2021}.
Through magnetization and specific heat measurements, we revealed a significant effective mass enhancement on approaching $H_{\rm m}$ \cite{Miyake2019,Imajo2019,AMiyake2021}.
A similar mass enhancement was also supported by resistivity measurements \cite{Knafo2019,Knafo2021} and explained  theoretically\cite{Miyake2021}. 
Note that above $H_{\rm m}$, the $b$ axis becomes the easy magnetization axis, while below $H_{\rm m}$, it is the $a$ axis with a quasi-saturation to $\sim$1~$\mu_{\rm B}$/U at $\mu_0 H \approx 20$~T \cite{Miyake2019,AMiyake2021}.
The singular point is that the slope $dM/dH$ for $H~||~b$ on both side of $H_{\rm m}$ is quasi constant, more precisely only $\sim$14\% reduction across $H_{\rm m}$, although the jump $\Delta M$ at $H_{\rm m}$ is generally regarded as a mark of drastic change in the localized 5$f$ contribution.

\begin{figure}
\begin{center}
\includegraphics[width=0.8\hsize]{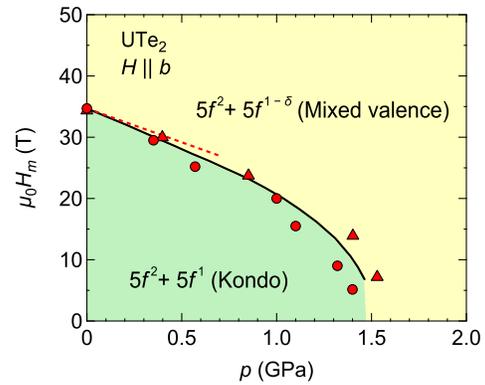}
\end{center}
\caption{(Color online)
Pressure dependence of $H_{\rm m}$ of UTe$_2$ for $H~||~b$.
The data points shown by circles and triangles are taken from Refs.~\citen{Knebel2020} and \citen{Lin2020}, respectively.
The dashed line indicates $\frac{\partial \mu_0H_{\rm m}}{\partial P} \approx -11.0$~T/GPa obtained by the Clausius-Clapeyron's equation.
The solid line is a guide for the eyes.
}
\label{PD}
\end{figure}

Another remarkable feature in UTe$_2$ is an intermediate valence state at ambient pressure revealed by uranium $L_3$ x-ray absorption near-edge spectroscopy (XANES) \cite{Thomas2020} and core-level photoelectron spectroscopy measurements\cite{Fujimori2021}. 
From the latter experiments, the dominant 5$f^3$ configuration is suggested to contribute to the uranium 5$f$ state in UTe$_2$. 
Moreover, the XANES results under pressure ($p$) revealed an apparent uranium valence shift of $\sim$0.1 across $p_c\approx 1.4$~GPa\cite{Thomas2020}, where the SC phase disappeared, instead, magnetically ordered phases appeard. \cite{Braithwaite2019,Ran2020,Aoki2020,Knebel2020,Lin2020,Aoki2021,Li2021}.
This valence shift coincides with a change in the magnetic anisotropy.
The easy magnetization axis switched from the $a$ to $b$ axes above $p_c$ \cite{Li2021}.  
At $p_c$, the $H_{\rm m}$ was also suppressed, as shown in Fig.~\ref{PD} \cite{Knebel2020,Lin2020}.
Thus, the valence shift was accompanied by the collapse of the metamagnetic transition at $H_{\rm m}$.
Such a valence instability is also expected at $H_{\rm m}$ at ambient pressure.
 
In this study, we measure the linear magnetostriction\cite{linearMS} of UTe$_2$ along all the principal axes for $H~||~b$ and estimate the resultant volume magnetostriction.
As expected from thermodynamic relations, a sharp step-like volume shrinkage across $H_{\rm m}$ is clearly observed.
Using the magnetostriction data, we discuss the metamagnetism from the thermodynamic point of view and a possible link to the valence instability at $H_{\rm m}$ at ambient pressure.

Single crystals of UTe$_2$ were grown using the chemical vapor transport method \cite{Aoki2019}.
The sample used here was cut from the same piece that reported the simultaneous measurements of magnetization and magnetocaloric effect (MCE) \cite{AMiyake2021}. 
Magnetostriction in pulsed-magnetic fields was measured using fiber-Bragg-grating (FBG) and the optical filter method \cite{Ikeda2018}.
The bare optical fibers with FBG were glued directly on the sample parallel and perpendicular to the field direction along the $b$ axis, as in Ref.~\citen{Gen2022}.
This procedure enabled us to measure the longitudinal and transverse magnetostriction simultaneously, the first attempt in a pulsed-field.
Thus, two datasets of linear magnetostriction along the $a$ and $b$ axes, and along the $b$ and $c$ axes, were measured to obtain the volume magnetostriction.
Almost identical results along the $b$ axis for the different experimental setups confirmed the reliability of our magnetostriction measurements.
The measurements were performed at low temperatures down to 1.4~K, where the sample was immersed in $^4$He liquid \cite{commentMCE}.
Pulsed-magnetic fields up to $\sim$54~T were generated using non-destructive pulse-magnets with typical pulses durations of $\sim$36~ms, installed at the ISSP of the University of Tokyo.

Figure \ref{MS}(b) shows the linear magnetostriction $\Delta L_i/L_i$ $(i = a, b, {\rm and}~c)$ and the volume magnetostriction $\Delta V/V$ at $T_{\rm ini}$ = 1.4~K for $H~||~b$.
In agreement with the previous study \cite{Thomas2021}, with increasing magnetic fields $\Delta L_c/L_c$ increases, while $\Delta L_a/L_a$ and $\Delta L_b/L_b$ decrease at low fields.
Discontinuous changes in the $\Delta L_i/L_i$s with the $H$-hysteresis were observed for all the axes, accompanied by the first-order metamagnetic transition at $H_{\rm m}$ [see also Fig.~\ref{MS}(a)]. 
Similar to the sign of the $dL_i/dH$ at low fields, an increase in $L_c$ and decreases in $L_a$ and $L_b$ across $H_{\rm m}$ for the field-up sweep were observed.
The largest discontinuity was observed for the $L_b$ as that in the low field region. 

First, we estimated the volume magnetostriction and compared it with the reported pressure measurements \cite{Knebel2020,Lin2020}. 
Discontinuous change in the volume across $H_{\rm m}$ at $T_{\rm ini}$ = 1.4~K is evaluated by $\Delta V/V = \sum_{i = a, b, c} (\Delta L_i/L_i)\approx -5.9\times10^{-4}$.
Adopting this value and $\Delta M\approx 0.5~\mu_{\rm B}$/U \cite{Miyake2019,AMiyake2021} to the Clausius-Clapeyron's equation, $\frac{\partial \mu_0H_{\rm m}}{\partial p} = \frac{\Delta V}{\Delta M}$, $\frac{\partial \mu_0H_{\rm m}}{\partial p} \approx -11.0~{\rm T/GPa}$ is obtained as the initial slope of $H_{\rm m}$ as a function of pressure.
Here, we used the reported molar volume $V_{\rm mol}\approx 5.22\times 10^{-5}~{\rm m^3/mol}$ at 2.7~K \cite{Hutanu2020}.
The direct comparison of this value to the $p$ dependence of $H_{\rm m}$ is shown in Fig.~\ref{PD}.
The $H_{\rm m}(p)$ determined thermodynamically and experimentally agrees very well \cite{commentCC}.
More details on the volume contraction at $H_{\rm m}$ will be discussed later.

\begin{figure}
\begin{center}
\includegraphics[width=\hsize]{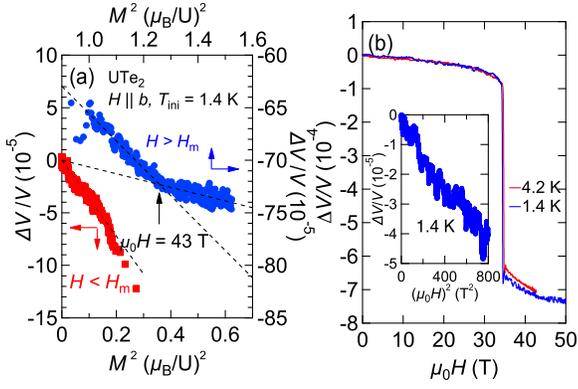}
\end{center}
\caption{(Color online)
(a) $M^2$ dependence of $\Delta V/V$ of UTe$_2$ at $T_{\rm ini}$~=~1.4~K for $H~||~b$.
The red squares (left and bottom axes) and blue circles (right and top axes) correspond to the data in field regions below and above $H_{\rm m}$, respectively. 
There is a kink structure at $\mu_0H = 43$~T, indicated by an arrow. 
The dashed lines indicate a relation, $\Delta V/V\propto M^2$.
(b) Field dependence of $\Delta V/V$ at $T_{\rm ini}$~=~1.4 and 4.2~K.
The inset shows $\Delta V/V$ as a function of the square of the fields at $T_{\rm ini}$~=~1.4~K.
}
\label{M2B_VMS}
\end{figure}

Using the magnetostriction and magnetization results, we can also derive the pressure dependence of the magnetic susceptibility by employing Maxwell's relation, $\left(\frac{\partial M}{\partial p}\right)_{H, T} = -\left(\frac{\partial V}{\partial \mu_0H}\right)_{p, T}$.
In a Pauli paramagnetic system, the magnetic susceptibility $\chi\equiv dM/d\mu_0H$ at low temperature is nearly constant.
Thus, the pressure derivative of the susceptibility is obtained as

\begin{equation}
\frac{d\chi}{dp}=-2C_vV_{\rm mol}\chi^2,
\label{dchidp}
\end{equation} 
where $C_v$ is the magnetovolume coupling constant, which is obtained by $\frac{\Delta V}{V} = C_vM^2$ \cite{Mushnikov2004}.
From the plot $\Delta V/ V$ versus $M^2$ below $H_{\rm m}$ at $T_{\rm ini}$~=~1.4~K shown as red squares in Fig.~\ref{M2B_VMS}(a), $C_v\approx -3.6\times10^{-4}~(\mu_{\rm B}{\rm /U})^{-2}$ was evaluated. 
Using the $C_v$ and $\chi$ for $H~||~b$ \cite{Miyake2019} to Eq.~(\ref{dchidp}), $\frac{d\chi}{dp}\approx 1.0\times 10^{-3}$~$\mu_{\rm B}$/(T$\cdot$GPa) is obtained.
This estimation agrees well with the reported pressure dependence of the magnetic susceptibility \cite{Li2021,comment_dchi}.
The proportionality of $\Delta V/ V \propto M^2$ holds even above $H_{\rm m}$, while a slope changes approximately at 43~T.
Although the origin of this anomaly is not clear at present, it may reflect the change in the 5$f$ characters.

As shown in Fig.~\ref{MS}(b), a strongly anisotropic magnetostriction (AMS) was observed.
Let us see the anisotropy $\Delta L_i^{\rm AMS}/L_i$ defined by $\frac{\Delta L_i^{\rm AMS}}{L_i}\equiv\frac{\Delta L_i}{L_i}-\frac{1}{3}\frac{\Delta V}{V}$\cite{Zieglowski_1986}.
The second term is an averaged lattice change expected for the isotropic case.
Figure \ref{MS}(c) depicts the AMS along each axis for $H~||~b$ at $T_{\rm ini}$ = 1.4~K.
$\Delta L^{\rm AMS}_a/L_a$ shows a weak field dependence, whereas $\Delta L^{\rm AMS}_b/L_b$ and $\Delta L^{\rm AMS}_c/L_c$ show significantly anisotropic behaviors. 
At $H_{\rm m}$, a discontinuous rise in $\Delta L^{\rm AMS}_c/L_c$ and discontinuous drop in $\Delta L^{\rm AMS}_b/L_b$ are more prominent.
Interestingly, the absolute values of the AMS for the $b$ and $c$ axes, including the discontinuous changes at $H_{\rm m}$, are almost identical.
These facts indicate the lattice instability in the $bc$ plane, which may be responsible for the field-induced SC phases around $H_{\rm m}$ with rotating the field direction between $b$ and $c$ axis \cite{Ran2019b,Knafo2021}.
The lattice instability in the $bc$ plane may also affect the antiferromagnetic  fluctuations with $k_1$~=~(0, 0.057, 0) \cite{Duan2020,Knafo2021b}.
Although the Fermi surface has not been clarified experimentally \cite{Fujimori2019, Miao2020}, the two-dimensionality is discussed from band calculations \cite{Ishizuka2019,Xu2019}.

To characterize the electronic state, \blue{Gr\"uneisen parameters of a characteristic energy $X$, $\Gamma_X\equiv-\frac{\partial \log{X}}{\partial \log{V}}$}, are useful \cite{Flouquet2005}.
Using the volume magnetostriction results below $H_{\rm m}$, the Gr\"uneisen parameters can be divided into magnetic ($\Gamma_H$) and thermal electronic ($\Gamma_T$) contributions \cite{Thalmeier1986, commentMagGru},

\begin{equation}
S_v =\frac{1}{2BV_{\rm mol}}\left[(2\Gamma_H-\Gamma_T)\chi+\Gamma_TT\chi'\right],
\label{Grun}
\end{equation}
where $S_v$ is a coefficient of the $H^2$-term of $\Delta V/V$, $B=57$~GPa the bulk modulus \cite{HondaPV}, and $\chi' = \partial\chi/\partial T$.
For $H~||~b$, more generally in Pauli paramagnets at low temperatures, there is less temperature dependence of $\chi$.
Thus, neglecting the second term of Eq.~(\ref{Grun}), we obtain $2\Gamma_H-\Gamma_T \approx -5$ from our magnetostriction  and magnetization results below $H_{\rm m}$ at $T_{\rm ini}$ = 1.4~K [see the inset of Fig.~\ref{M2B_VMS}(b)]: $S_v = -5.6\times 10^{-8}~{\rm T}^{-2}$ and $\chi\approx 0.0122~{\mu_{\rm B}}/{\rm T}$.
Using the reported $\Gamma_T\approx -30$ \cite{Thomas2021,Willa2021,Li2021}, $\Gamma_H\approx-17.5$ is derived. 
$\Gamma_H$ is nearly half of $\Gamma_T$.

\begin{figure}
\begin{center}
\includegraphics[width=0.8\hsize]{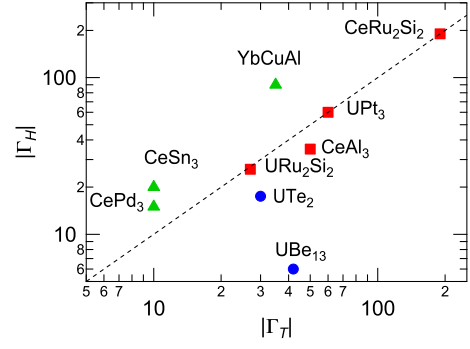}
\end{center}
\caption{(Color online)
Comparison between absolute values of the thermal ($\Gamma_T$) and magnetic ($\Gamma_H$) Gr\"uneisen parameters in various compounds.  
The broken line indicates $|\Gamma_T| = |\Gamma_H|$.
As plotted in different symbols, these intermetallic compounds may be classified into intermediate valence systems ($|\Gamma_T| <|\Gamma_H|$, triangles), heavy fermion systems ($|\Gamma_T| \approx|\Gamma_H|$, squares), and the others ($|\Gamma_T| > |\Gamma_H|$, circles).
The data except for this study (UTe$_2$) are obtained from Refs.~\citen{Kaiser1988, deVisser1992, Dijk1995, Lacerda1989}.
}
\label{Gru}
\end{figure}

Figure \ref{Gru} compares the relation of the absolute values of $\Gamma_T$ and $\Gamma_H$ in various compounds \cite{Kaiser1988, deVisser1992, Dijk1995, Lacerda1989}.
For many heavy fermion systems, $\Gamma_T = \Gamma_H$ is held \cite{Flouquet2005,Kaiser1988}.
Thus, only a single characteristic energy scale dominates the electronic and the magnetic properties in many heavy fermion systems.
In other words, a dominant interaction governs the electronic properties in many heavy fermion systems.
Deviations from the correspondence may be seen in the intermediate valence systems, such as CeSn$_3$, CePd$_3$, and YbCuAl, whose $|\Gamma_H|$ is larger than $|\Gamma_T|$.
Another example of $|\Gamma_T| > |\Gamma_H|$ was reported for unconventional superconductor UBe$_{13}$\cite{deVisser1992}, where it is well known that the band structure and non-Fermi liquid state are strongly-field dependent \cite{Shimizu2015,comment_UBe13}. 
A sole magnetic property may not link the electronic property.
In UTe$_2$, $\Gamma_T$ is nearly double of $\Gamma_H$, indicating different channels of interactions. 

From early experimental investigations, the link between $T_{\chi}^{\rm max}\approx 35$~K and $\mu_0H_{\rm m}\approx35$~T was discussed \cite{Knafo2019,Miyake2019,Knebel2020,AMiyake2021,Li2021}, as in other paramagnetic systems \cite{Aoki2013}.
Recently, a different characteristic energy scale $T^{\ast}\approx 12$~K was also proposed \cite{Niu2020,Cairns2020,Eo2021,Willa2021,Tokunaga2022}.
A broad peak-like anomaly appeared at $T^{\ast}$ in some thermodynamic quantities\cite{Willa2021}.
NMR studies for $H~||~a$ have revealed some anomalies at both of $T_{\chi}^{\rm max}$ and $T^{\ast}$; starting a broadening of NMR spectra below $T_{\chi}^{\rm max}$ and a peak in the nuclear spin-spin relaxation rate at $T^{\ast}$ \cite{Tokunaga2022}.
Thus, $T^{\ast}$ may correspond to the short-range ordering temperature. 
Interestingly, $\Gamma_H$ is almost identical to $\Gamma_T$ at $T^{\ast}$ \cite{Willa2021}.
This coincidence indicates that the low field short-range order is eliminated at $H_{\rm m}$.

To further address the influence of the metamagnetic transition on the electronic properties, we discuss $\Gamma_T$ above $H_{\rm m}$.
We estimate the volume thermal expansion using magnetostriction results at $T_{\rm ini}$ = 1.4 and 4.2~K, as shown in Fig.~\ref{M2B_VMS}(b).
Assuming the isothermal process above $H_{\rm m}$, the volume thermal expansion coefficient $\alpha_V\equiv V^{-1}dV/dT\sim [\Delta V(4.2~{\rm K})/V-\Delta V(1.4~{\rm K})/V]/(4.2 -1.4)\approx$ $7\times 10^{-6}$~K$^{-1}$ was estimated at 40~T, slightly higher than $H_{\rm m}$.
Using the field dependence of the electronic specific-heat coefficient obtained by our previous measurements \cite{Imajo2019,AMiyake2021}, $\Gamma_T \approx$ 110 at 40~T was obtained\cite{commentGru}.
The derived large positive value of $\Gamma_T$ above $H_{\rm m}$ is a mark that the Sommerfeld coefficient $\gamma\propto C/T$ has a strong maximum at $H_{\rm m}$ \cite{Miyake2019,Imajo2019,AMiyake2021}.

The volume contraction in UTe$_2$ at $H_{\rm m}$ for $H~||~b$ is strikingly different from the typical example of heavy-fermion pseudo-metamagnetic transitions, such as CeRu$_2$Si$_2$ \cite{Lacerda1989} and UPt$_3$ \cite{deVisser1987}, which show an increment in volume at $H_{\rm m}$.
Note that the decrease of $\Delta L_b/L_b$ suggests that the uniaxial pressure for the $b$ axis will reduce $H_{\rm m}$. 
This is indeed observed at the spin reorientation field $H_R$ in URhGe \cite{Braithwaite2018,Nakamura2018}.
The volume contraction of UTe$_2$ at $H_{\rm m}$ may correspond to the loose of the itinerant character and the resultant valence change.

We discuss a possible link of the metamagnetic transition to the valence instability. 
As shown in Fig.~\ref{PD}, $H_{\rm m}$ is suppressed at $p_c$.
Increasing pressure from the intermediate valence state (U$^{3+}$ and U$^{4+}$) at $p=0$, the uranium valence increases $\sim$0.1 at $p_c$, leading the system to a more tetravalent 5$f^2$ configuration \cite{Thomas2020,Fujimori2021}.
Thus, a volume contraction accompanied by the valence change at $p_c$ is expected.
Such a contraction is indeed observed across $H_{\rm m}$ at ambient pressure.
A significant change is expected to occur in the 5$f$ uranium magnetism as functions of pressure and magnetic field.
 
The observed metamagnetic behaviors can be naturally understood on the basis of the so-called itinerant-localized duality model for the $5f^3$-based heavy fermion system, in which the $5f^2$-electron state with the less hybridization between conduction and $f$ electrons behaves as $localized$ with the $5f^2$ crystalline electric field level scheme, while the $5f^1$-electron state with the larger hybridization behaves as $itinerant$. 
Indeed, in the case of UPd$_2$Al$_3$ (a $5f^3$-based heavy fermion system), this physical picture was shown to work quite well \cite{Sato2001}. 
The crucial point is that the $itinerant$ $5f^1$ component behaves similarly to quasiparticles of $4f^1$-based heavy fermion systems, that is, Ce-based system. 
UTe$_2$, in the nearly $5f^3$ configuration at an ambient condition, would  also exhibit the duality of the $5f$ electrons, as in UPd$_2$Al$_3$. 
Particularly, the component of electrons in the $5f^1$-electron state can exhibit a discontinuous metamagnetic transition associated with a first-order valence transition from Kondo to mixed valence (MV) state as discussed in Refs. \citen{Watanabe2008} and \citen{Watanabe2009} (see also the scheme in Fig.~\ref{PD}). 
In this case, in the region below $H_{\rm m}$, 5$f^1$ components are in the so-called Kondo state (the $f$ electron number $n_f\simeq$ 1), but in the region above $H_{\rm m}$, 5$f^1$ components are in the MV (or valence fluctuating) state. 
This type of behavior seems to reproduce the magnetic field and pressure dependence of the valence of 5$f$ electrons observed in UTe$_2$ across the metamagnetic transition, as shown in Fig~\ref{PD}.
Thus, the electronic correlations decrease drastically with feedback on the band structure at $H_{\rm m}$. 
This picture seems to be supported by the large jump of the magnetoresistance \cite{Knafo2019,Knafo2021}, the reduction of carrier density \cite{Niu2020b}, and the drop of $\gamma$ \cite{Imajo2019,AMiyake2021}.
The corresponding changes in Fermi surface and magnetic interactions are ingredients of metamagnetism.
Although the link of valence transition across $H_{\rm m}$ at $p=0$ and across $p_c$ at $H=0$ is not much evident, the same trend of magnetization-easy-axis switch from the $a$ to the $b$ axes across $H_{\rm m}$ and $p_c$ were observed. \cite{Miyake2019,AMiyake2021,Li2021}. 
Qualitatively it seems a promising road; quantitatively theoretical progress is now necessary.  

In summary, we demonstrated the magnetostriction measurements of UTe$_2$ for $H~||~b$.
A clear drop in the volume magnetostriction accompanied by the first-order metamagnetic transition was observed, which satisfactorily agrees with the pressure dependence of $H_{\rm m}$.
Significant anisotropic linear magnetostriction was also revealed; that for the $a$ axis is almost identical to the averaged volume magnetostriction, while those for the $b$ and $c$ axes show larger magnetostriction with a similar magnitude but with opposite sign.
This anisotropy indicates lattice instability within the $bc$ plane, which may trigger the field-reinforced/reentrant superconductivity. 
Through the Gr\"uneisen parameter analyses, we discussed the relation between characteristic energy scales in temperatures and fields.
A possible link of the metamagnetic transition to a valence instability is considerable on the basis of the dual nature of the itinerant-localized 5$f$ electrons. 
This suggests the low- and high-field electronic and magnetic properties are crucially governed by different interactions.
The magnetic field dependence of the duality deserved to be clarified.

\begin{acknowledgment}
We thank G. Knebel and W. Knafo for fruitful discussions.
This work was supported by KAKENHI (JP15H05884, JP15H05882, JP15K21732, JP16H04006, JP15H05745, JP17K05555, JP19H00646, JP20K03854, JP20K20889, JP20J10988, JP20H00130, and JP20KK0061), ICC-IMR, and ERC starting grant (NewHeavyFermion).
\end{acknowledgment}

\newpage

\section*{Supplemental material}

\section{Comparison of $\chi(p)$ determined thermodynamically and experimentally}

\begin{figure}[h]
\begin{center}
\includegraphics[width=1\hsize]{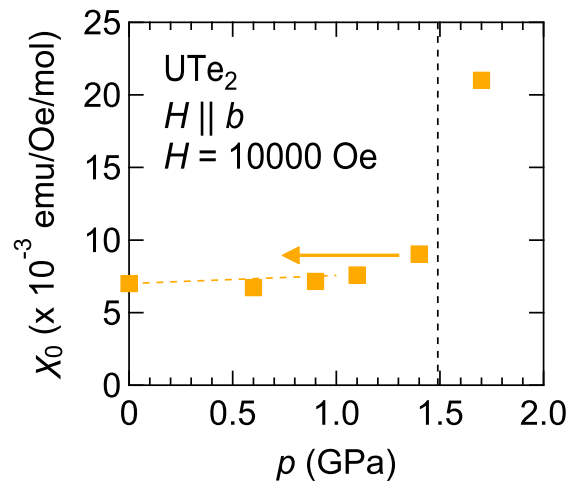}
\end{center}
\caption{Pressure dependence of $\chi_0$ at 10000~Oe for $H~||~b$ of UTe$_2$.
The dashed line indicates $d\chi/dp\approx 5.6\times10^{-4}$~emu/(mol$\cdot$ Oe$\cdot$ GPa).
The vertical dashed line indicates a critical pressure $p_c$.
The $\chi_0(p)$ data are taken from Ref.~\citen{Li2021}.
}
\label{chiP}
\end{figure}

We compare the pressure dependence of the magnetic susceptibility $\chi$ for $H~||~b$ determined thermodynamically and experimentally. 
Figure~\ref{chiP} shows pressure dependence of $\chi_0$ at 10000~Oe for $H~||~b$, which is estimated by extrapolating to 0~K \cite{Li2021}. 
From Maxwell's relation using our magnetization and magnetostriction results, $\frac{d\chi}{dp}=5.6\times 10^{-4}$~emu/(mol$\cdot$ Oe$\cdot$ GPa) is obtained.
As shown by a dashed line in Fig.~\ref{chiP}, our evaluated $\chi_0(p)$  agrees nicely with pressure dependence of $\chi_0$.

\begin{thebibliography}{99}
\bibitem{linearMS} The idiom ``linear magnetostriction''  used in this manuscript does not mean the magnetostriction in proportion to magnetic fields due to piezomagnetism.
We use it to distinguish it from ``volume magnetostriction''.
\bibitem{commentMCE} Since the MCE accompanied by the metamagnetic transition is not negligible, even the sample is immersed in the superfluid $^4$He.
The sample temperature above $H_{\rm m}$ changes. \cite{AMiyake2021}
Thus, $T_{\rm ini}$ in this manuscript is the temperatures at the beginning of the field generation.
\bibitem{commentGru} Here, we follow the conventional definition of the Gr\"uneisen parameter, $\Gamma_T = \frac{V_{\rm mol}\alpha_V}{\kappa C}$, where $\kappa=1/B$ and $C$ are compressibility and specific heat, respectively.
\end{thebibliography}
\end{document}